# Cooperative Multi-Agent Deep Reinforcement Learning for Adaptive Decentralized Emergency Voltage Control

Ying Zhang, *Member, IEEE*
Department of Electrical and Computer Engineering
Montana State University
Bozeman, MT, U.S.

Meng Yue, *Member, IEEE*
Interdisciplinary Science Department
Brookhaven National Laboratory
Upton, NY, U.S.

*Abstract*—Under voltage load shedding (UVLS) for power grid emergency control builds the last defensive perimeter to prevent cascade outages and blackouts in case of contingencies. This letter proposes a novel cooperative multi-agent deep reinforcement learning (MADRL)-based UVLS algorithm in an adaptive decentralized way. With well-designed input signals reflecting the voltage deviation, newly structured neural networks are developed as intelligent agents to obtain control actions and their probabilities to accommodate high uncertainties in volatile power system operations. Moreover, the interaction among the agents for coordinated control is implemented and refined by a state-of-the-art attention mechanism, which helps agents concentratively learn effective interacted information. The proposed method realizes decentralized coordinated control, adapting to extremely high uncertainties. Case studies on an IEEE benchmark system indicate the superior performance of the proposed algorithm.

## I. Introduction

AFTER occurrences of contingencies, power grid emergency control is vital to reduce the chance of occurrence and impact of power outages. Under voltage load shedding (UVLS) is the last resort yet one of the most effective alternatives widely adopted for real-world emergency control [1], falling into the category of short-term voltage stability (SVS) control. Moreover, UVLS requires mitigating fault-induced delayed voltage recovery (FIDVR) events, which are the phenomena whereby bus voltages drop rapidly to significantly reduced levels after fault clearance. Traditional UVLS schemes are usually rule-based and designed offline, in which a fixed number of loads at individual buses are shed if the bus voltages are observed to fall below pre-set thresholds over a certain amount of time [2]. The *if-then* mode without awareness of the future voltage trajectories, however, makes it difficult to recover voltages timely.

Afterward, various online UVLS schemes are explored, such as model predictive control (MPC) [3], fuzzy logic [4], and extreme learning machines [5]. Built on a centralized architecture, these approaches depend on a central controller. Decentralized control schemes can mitigate communication time delay of observation data and accelerate data acquisition procedures. Consequently, the control reliability and responsiveness are improved significantly, e.g., see [6]–[8] and the references therein. [7] develops a decentralized multi-agent UVLS method, which coordinates the agents by identifying the severity of disturbance in each area individually. However, [6]–[8] are model-based, and their control performance is heavily subject to modeling complexity. Moreover, due to the velocity of power system dynamic evolution, the model-based methods envisage huge computational burdens to predict dynamic system behaviors and determine the UVLS strategies.

On the other hand, the methods [3]-[8] cannot adapt to highly uncertain fault scenarios, although they can provide satisfactory results for some presumed cases. Deep reinforcement learning (DRL) is recently developed to cater to increasing uncertainties and complexities in power systems, e.g., [9], [10] for UVLS and [11] for circuit linearization. [10] develops a deep Q-network (DQN) algorithm to implement adaptive load shedding in a known heavily loaded area. However, the action-state space will exponentially expand when coming to multi-area power systems with an uncertain selection of shedding areas. The DQN-based method suffers from the "curse of dimensionality" issue [12] and thus easily produces regressive performances in the high-dimensional action-state space. [9]–[11] run in a centralized fashion, while multi-agent DRL (MADRL) is able to offer decentralized solutions to system decision-making, e.g., [13]. Yet, adaptive decentralized UVLS is largely missing. The drastic difficulties lie in solving the coordinated control variables from local observation while adapting to the volatile operation of uncertain fault-triggered dynamic power systems. It should be noted that three-aspect uncertainties exist in practical power systems and should be reckoned with, including high uncertainties from 1) the combination of varying operating conditions and fault scenarios, 2) sensitive dynamic circuit evolution of large-scale differential algebraic equations (DAEs), and 3) varying voltage recovery requirements (as we will elaborate in Section II). These high uncertainties make shedding commands easily inappropriate, and the short-term voltage instability issue, which can be catastrophic, might arise. Therefore, it is desired to design a MADRL framework with high adaptivity to such multi-aspect uncertainties and realize coordinated decentralized control simultaneously.

This paper proposes a novel cooperative MADRL-based UVLS algorithm for decentralized emergency SVS control to adapt to high uncertainties. We model the controller for each subnetwork as a deep neural network (NN)-based agent and restructure the agent to learn the discrete probabilities of feasible actions in the highly uncertain environment. Moreover, our approach embeds an attention mechanism to facilitate

This work was supported by the Advanced Grid Modeling Program, Office of Electricity Delivery and Energy Reliability of the U.S. Department of Energy.

multi-agent cooperative learning of the control policy and automatically selects which agents to attend to the emergency control adequately. In an online implementation, the proposed method achieves highly adaptive decentralized control by only using local observation for execution. The main contributions are listed below.

1) We develop an attention-embedded mechanism to derive automatic weights of information interaction for different areas impacted by unknown contingencies. This yields adaptive, non-iterative selection of the most responsive shedding areas.

2) The well-designed NNs enable probability estimation of discrete UVLS actions, and thus the proposed algorithm has high multi-aspect adaptivity to extremely uncertain operating environments.

3) Comprehensive case study demonstrates that the proposed method achieves faster convergence and more efficient and scalable learning than several existing approaches.

## II. PROBLEM DESCRIPTION FOR UVLS AGAINST FIDVR

The challenge of efficient UVLS schemes roots in handling the set of DAEs to perceive the dynamic behaviors of a power system triggered by faults. Moreover, UVLS for emergency control in the system is a highly nonlinear and non-convex constrained optimization problem, generally formulated as [3]:

$$\min_{\boldsymbol{u}} C(\boldsymbol{u}(t)) \tag{1}$$

$$\text{s. t.} \begin{cases} \dot{\boldsymbol{x}}(t) = \Psi(\boldsymbol{x}(t), \boldsymbol{y}(t), \boldsymbol{u}(t)) & (1a) \\ 0 = \Phi(\boldsymbol{x}(t), \boldsymbol{y}(t), \boldsymbol{u}(t)) & (1b) \\ \underline{\boldsymbol{S}} \leq G(\boldsymbol{x}(t), \boldsymbol{y}(t), \boldsymbol{u}(t)) \leq \overline{\boldsymbol{S}} & (1c) \\ \underline{\boldsymbol{u}} \leq \boldsymbol{u}(t) \leq \overline{\boldsymbol{u}} & (1d) \end{cases}$$

where $t \in [t_0, t_0 + T]$, $T$ is the control horizon time, and $t_0$ denotes the control beginning time; $\boldsymbol{x}$ is a vector of state variables, such as rotor angles and angular speeds; $\boldsymbol{y}$ is an algebraic state vector of the power grid, which are typically the voltages at buses of the power system; $\boldsymbol{u}$ is a vector of control variables (actions), and for UVLS, it denotes the shedding loads at all the controllable buses; $\Psi(\cdot)$ denotes differential functions to describe system state equations, and $\Phi(\cdot)$ denotes algebraic functions for network operation; $C(\boldsymbol{u}(t))$ is a control cost function. During the procedure, the control strategy is implemented per $T_c$. Note that, the network equations involve the operations of the system network with a fault clearance.

**Remark.** Due to the page limit, the modeling details about the dynamic circuit components used for emergency voltage control are omitted here and can be found in [3] and [9].

Furthermore, against FIDVR events, the transient voltage recovery criteria (TVRC) are widely used during the UVLS procedure [10], [14] and thus adopted here. Per these criteria, the voltages are required to restore to at least 0.7, 0.8, 0.9, and 0.95 levels of the nominal values within 0, 0.33, 0.5, and 1.5 s, respectively, after a fault is cleared at $T_{fc}$. Please see the typical TVRC envelope in [14].

Given $N_L$ controllable nodes in the power system, the UVLS actions at $t$ that are taken at the controllable nodes are expressed as

$$\boldsymbol{u}(t) = [u_1, u_2, \ldots, u_{N_L}] \tag{2}$$

The objective function for UVLS is to minimize the total load shedding costs in all the controllable nodes in the system at $t$:

$$C(\boldsymbol{u}(t)) = \sum_{i=1}^{N_L} P_{Li} u_i \tag{3}$$

where $P_{Li}$ denotes the initial load at the $i$th controllable nodes.

## III. PROPOSED MADRL-BASED ALGORITHM

The proposed algorithm enables NN agents to learn the probabilities of feasible discrete actions, resulting in a highly efficient DRL structure. Upon this structure, we construct an attention mechanism working among the agents to identify the severity of disturbance in a non-iterative intelligent fashion. This attention-embedded MADRL algorithm realizes coordinated decentralized voltage control with low shedding costs efficiently in power systems.

Post-fault multi-area power system acts as the DRL environment. In the multi-agent architecture, an agent is assigned to one of the areas [7]. The proposed MADRL-based UVLS method interprets the voltages, the control variables, and the objective function in (4) into the state, action, and reward in the agents, respectively. They are briefly explained below.

1) *Action:* At time $t$, the control action in agent $j$ is defined as $\boldsymbol{a}_j^t = \boldsymbol{u}_j(t)$, where $\boldsymbol{u}_j \subseteq \boldsymbol{u}$ is the discrete action at the controlled load bus(es) in the agent [7], and $j \in \{1, 2, \ldots, N\}$.

2) *State:* The existing DRL-based UVLS methods, e.g., [10], tend to use direct voltage observation as the state. The UVLS problem against fast-varying FIDVR is devoted to predicting system voltage violations from the time-varying TVRC requirements rather than directly inferring system voltage trajectories. To directly inform these requirements of the agents, we propose taking advantage of the voltage deviation from the TVRC to be the input of the DRL agents, *i.e.*, the state. The latest $N_r$ voltage deviations in an area are collected as the states of agent $j$: $\boldsymbol{s}_j^t = [\boldsymbol{O}_{t-N_r-1}, \ldots, \boldsymbol{O}_{t-1}]$, where $\boldsymbol{O}_\tau = \{\Delta \boldsymbol{V}_j(\tau)\}$ and $\tau \in \{t - N_r - 1, \ldots, t - 1\}$, and the deviations on bus $i$ in the area are computed by

$$\Delta V_i(t) = \begin{cases} V_i(t) - 0.7 & t \in [T_{fc}, T_{fc} + 0.33) \\ V_i(t) - 0.8 & t \in [T_{fc} + 0.33, T_{fc} + 0.5) \\ V_i(t) - 0.9 & t \in [T_{fc} + 0.5, T_{fc} + 1.5) \\ V_i(t) - 0.95 & t \in [T_{fc} + 1.5, t_0 + T] \end{cases} \tag{4}$$

3) *Reward:* The reward at time $t$ is calculated for agent $j$ as:

$$r_j^t = \begin{cases} -M & \text{If } \forall i, \Delta V_i(t) < 0, \text{ calc. by (4)} \\ P_{Lj}(1 - u_j(t)) & \text{Otherwise} \end{cases} \tag{5}$$

where $M$ denotes a penalty when TVRC constraints on one or more buses are violated; the remaining loads are computed in terms of percentages of the initial controllable loads.

### B. DRL Structure for UVLS

The existing DQN-based algorithm [10] only learns deterministic actions by finding the maximum Q-value corresponding to those actions, resulting in poor efficiency in a volatile environment. The conventional SAC method is subject to search continuous action space [12] and cannot be directly applied to the dedicated UVLS problem. To this end, updating the traditional DRL structure for providing discrete UVLS actions adapting to such an environment is non-trivial. Therefore, we propose a discrete-actor-critic DRL structure to learn the discrete probability distribution of the UVLS actions.

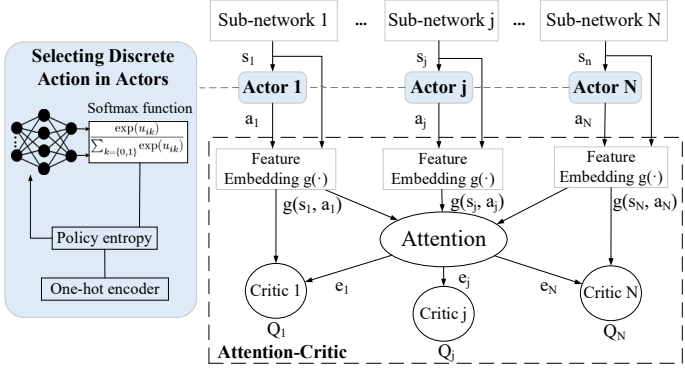

Fig.1. Proposed MADRL framework and actor structure.

The proposed DRL structure contains three types of NNs, i.e., a soft Q-function $Q$ parameterized by $\theta$ and a value function for policy evaluation, and a policy function $\pi$ parameterized by $\phi$ for the discrete policy update. The objective function of the proposed method is to maximize the sum of the expected reward and policy entropy of the discrete shedding actions for high adaptivity.

The actor network for the UVLS problem learns a stochastic map from states to actions, which can handle the very high uncertainties from varying pre-fault operating conditions and fault scenarios. The actor $\pi_\phi^j(s_j)$ in agent $j$ learns the localized policy based on its individual state $s_j$. To offer the discrete actions, we attach a softmax function in the last layer of the actor NN to estimate the probability of each feasible action, as shown in Fig.1. Then, the one-hot encoding technique is used to calculate the policy entropy $\mathcal{H}(\cdot|s_j) = -\alpha \log(\pi(\cdot|s_j))$.

We adopt a soft Bellman equation integrating the policy entropy, expressed as [12]

$$Q_j(s_j, a_j) = r_j + \gamma \mathbb{E}_{(s_j,a_j)\sim\rho_\pi}[Q_j(s'_j, a'_j) - \alpha \log(\pi(a_j|s_j))] \quad (6)$$

where $\gamma$ is a discount factor, and $\alpha$ is a temperature parameter.

The soft Q network used as the critic for evaluating the policy is trained by minimizing the soft Bellman squared residual for $(s_j, a_j)$ pairs, which are sampled from in a replay buffer.

Next, we further propose an attention-embedded MADRL algorithm to realize cooperative learning among the agents.

*C. Attention-enabled MADRL Algorithm*

Based on the above DRL structure, the proposed MADRL method is shown in Fig.1 and consists of discrete actors and attention-critics for each subnetwork.

Attention [15] is an NN-based computation system originally proposed for natural language processing (NLP). In this paper, we strategically embed the attention mechanism in the agents to identify the most affected area by uncertain contingencies randomly happening in the grid. For policy evaluation, the attention-critics selectively concentrates on other agents' contributions and eliminates distraction from irrelevant ones to evaluate the Q-value more efficiently.

The actor in an agent performs the policy update individually based on this evaluation to provide the distribution of proper actions. The attention-critic in agent $j$ adopts a NN $f_j$ to approximate the Q-value function for the policy evaluation, and the impacts of other agents are quantified by:

$$Q_j(s_j, a_j) = f_j(g_j(s_j, a_j), e_j) \quad (7)$$

where $g_j(s_j, a_j)$ is a multi-layer perceptron feature embedding function for the low-dimensional representation; $e_j$ is a weighted sum of the Q function outputs from other agents, implying their contribution to agent $j$ and is obtained by the attention system.

In the attention mechanism, the latent feature representation from the agents is assembled and used for cooperative Q learning. Specifically, the attention system is integrated into the original critic and computes a weighted sum of the Q outputs based on query-key-value tuples $(q_j, K_j, V_j)$ in agent $j$. The query, key, and value are computed by multiplying $g_j(s_j, a_j)$ by transformation matrices, $W_q$, $W_k$, and $W_v$.

The key $V_j$ is a function of the embedding, expressed as:

$$V_j = W_v g_j(s_j, a_j) \quad (8)$$

The weighted sum of the Q function outputs from other agents $e_j$ is calculated as:

$$e_j = \sum_{i\in\{1,2,\ldots N\}\setminus j} \alpha_i \, \text{LReLU}(V_i) \quad (9)$$

where a leaky Rectified linear unit as an activation function $\text{LReLU}(\cdot)$ is added to improve the efficiency; $\alpha_j$ is acquired by comparing the similarity between the embedding of agent $j$ and other agent $i$; specifically, a softmax function is used to quantify this similarity on the value [16]:

$$\alpha_j \propto \exp\left(g_j(s_j, a_j)^\mathrm{T} W_k^\mathrm{T} W_q g_i(s_i, a_i)\right) \quad (10)$$

The attention-critic $Q_j$ is jointly parameterized by the parameters of the critic function $\theta_j$ and $\{W_k, W_q, W_v\}$. These parameters in all the attention-critics are updated together by minimizing the following regression loss function:

$$\mathcal{L}(\theta) = \sum_{i=1}^{N} \mathbb{E}_{(s,a,r,s')\sim D}[(Q_i(s_i, a_i) - y_i)^2] \quad (11)$$

$$y_i = r_i + \gamma \mathbb{E}_{a'\sim\pi_{\bar\phi}(s')}\left[\left(Q_i^{\bar\theta}(s'_i, a'_i) - \alpha \log(\pi_{\bar\phi}(a_i|s_i))\right)^2\right] \quad (12)$$

where $Q_i^{\bar\theta}(s'_i, a'_i)$ is the target attention-critic agent, and $\bar\theta$ and $\bar\phi$ denote the NN parameters of the target attention-critic and target policy; $D$ is a replay buffer.

Then the policy gradient [12] is used for the actor's policy update to optimize the action in agent $j$ by

$$\widehat{\nabla}_{\phi_j}\mathcal{L}_\pi(\phi_j) = \mathbb{E}_{s_t, a_t \sim \pi(\phi)}\left[p(s_t, a_t)\widehat{\nabla}_{a_j} Q_j(s_j, a_j)|_{a_j=p(s_j)}\right] \quad (13)$$

## IV. CASE STUDY

We test the proposed algorithm on the IEEE 16-machine 68-bus power system [17]. This benchmark system is abstracted from the real-world interconnected New England test system (NETS) with New York power system (NYPS) and simulated in Power System Toolbox (PST) [18], which is widely used for dynamic system time-domain simulation. To sufficiently simulate the system dynamics, various components, such as sub-transient generators, thermal turbine governors, static exciters, load models, etc., are used [19]. This system is divided into six areas, as shown in Fig.2.

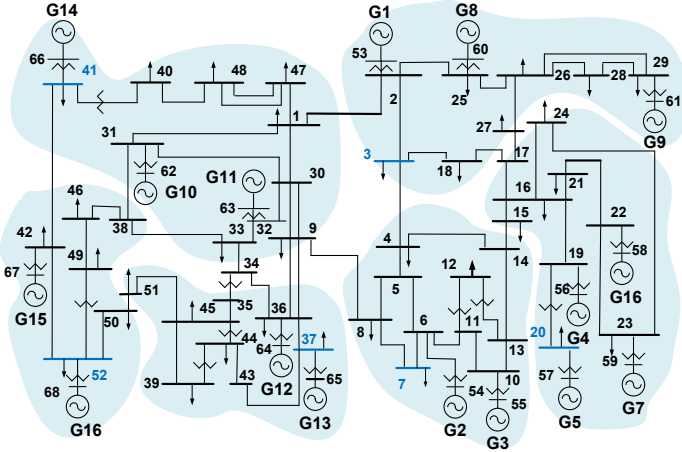

Fig. 2. The IEEE 16-machine 68-bus power system.

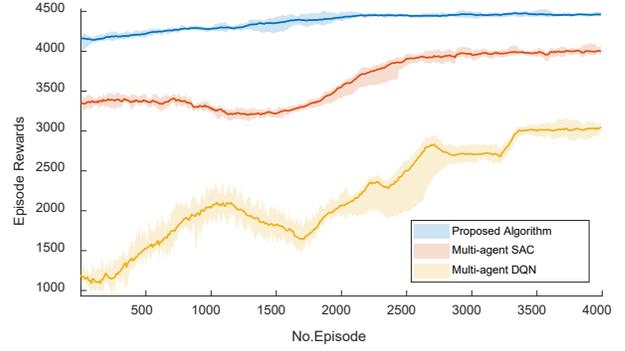

Fig.3. Comparison of average moving rewards in offline training.

TABLE I
PARAMETER SETTING IN THE PROPOSED METHOD

| DRL Parameters | Value | DRL Parameters | Value |
|---|---|---|---|
| Number of hidden layers | 2 | Discount factor $\gamma$ | 0.99 |
| Size in each hidden layer | 128 | Replay buffer size | 8000 |
| Learning rates for actors | 0.0001 | Mini-batch size | 3072 |
| Learning rates for critics | 0.0001 | Smoothing para. $\tau$ | 0.005 |

TABLE II COMPARISON OF OFFLINE TRAINING PERFORMANCE IN DIFFERENT ALGORITHMS

| MADRL Algorithms | Offline Training | | Online Test |
|---|---|---|---|
| | No. Episode of convergence | Average reward after 3500 episodes | Average reward |
| MA-DQN | about 2151 | 3063.1 | 2981.5 |
| MA-SAC | about 2911 | 3971.3 | 3983.2 |
| **Proposed** | about 3392 | 4463.9 | 4367.6 |

To investigate the adaptivity of the proposed UVLS method, various fault scenarios are set. They are randomly generated by combining different load variations (90% to 120% of the base values [10]), fault types, fault duration time (floating in [0.06s, 0.1s]), and locations of fault lines, yielding incredibly high variations. The controllable loads are located on buses 3, 7, 20, 37, 41, and 52. Here we test a five-round UVLS scheme, i.e., the number of action steps is 5 [3]. At each action time step, the control agents determine whether and which of the controllable nodes to shed 10% of the initial load [10], and the shedding actions are taken with a control interval of per second. The sampling time interval is 0.1 seconds. The NN parameters are listed in Table I. In the online test, 1,000 new random scenarios are performed for the evaluation of online control.

*1) Offline Training Performance.* For comparative study, we execute multi-agent DQN and SAC (MA-DQN and MA-SAC) methods. For a fair comparison, here we update the conventional SAC [12] by using the proposed NN design in Section III-B for probability learning. The moving average rewards of these algorithms in the offline training are shown in Fig. 3, and the average rewards and convergency performance of these algorithms are summarized in Table II. It can be observed that the proposed method obtains the highest reward, since the combination of the attention mechanism and the discrete-actor-critic DRL structure strategically facilitates the agents to extract more coordinated control strategies. Furthermore, the offline training of the proposed method is observed to converge at about 2,151 episodes, at the fastest convergence speed among all these methods. As shown in Table II, the average rewards in the 1000 online tests are 4,367.6 in the proposed method vs. 3,983.2 in the MA-SAC algorithm. The higher average reward, in both offline training and online test, implies improved voltage recovery performance and low load shedding cost.

*2) Overall Performance Evaluation.* To comprehensively evaluate the performances of the proposed algorithm on online UVLS, we adopt four performance metrics for statistical measures in $N_{test}$ random new cases for online UVLS tests. Two of them are the average percentage of load curtailment to the initial loads $\bar{P}_{dev}$ and the failure rate for quantifying the probability of unstable cases $R_{fal}$, in which voltage instability occurs due to inappropriate actions. $\bar{P}_{dev}$ and $R_{fal}$ are defined in [19]. Also, to assess the satisfaction of TVRC that imposes precise voltage recovery requirements at several pre-set time snapshots, here we investigate two additional indices to complement, which are computed by

$$\text{Mean voltage deviation } \bar{V}_{dev} = \frac{1}{n}\sum_{i=1}^{n} \Delta V_i(t_{\text{end}}) \quad (14)$$

$$\text{TVRC success rate } R_{TVRC} = N_{tvrc}/N_{test} \times 100\% \quad (15)$$

where $\bar{V}_{dev}$ denotes mean voltage deviation at all the buses w.r.t. the TVRC envelope after all the time-series actions are executed, and a positive $\bar{V}_{dev}$ indicates that the TVRC constraints are satisfied according to (5); $N_{tvrc}$ denotes the number of successful cases in terms of voltage recovery, and a successful case here is defined as the case where no voltage violation from TVRC occurs during the whole control process. The higher $R_{TVRC}$ and lower positive $\bar{V}_{dev}$ values jointly imply efficient control strategies that can timely meet and get closer to the TVRC envelope, while the lower $\bar{P}_{dev}$ demonstrates efficient control with lower shedding costs.

Given $N_{test} = 1000$, we compare these indices of different methods, which are listed in Table III. Apart from the mentioned DRL algorithms, a five-round rule-based UVLS scheme used in [19] is also compared, which is implemented by checking if the bus voltages satisfy a series of TVRC rules [2].

TABLE III COMPARISON OF PERFORMANCE METRICS IN ONLINE TEST
(1000 NEW CASES)

| Methods | $R_{fal}$ [%] | $\bar{P}_{dev}$[%] | $\bar{V}_{dev}$ [pu] | $R_{TVRC}$[%] |
|---|---|---|---|---|
| No control | 28.31 | - | -0.1883 | 0 |
| Rule-based [19] | 0 | 18.76 | 0.1743 | 12.43 |
| MA-DQN | 11.76 | 16.78 | 0.1524 | 71.64 |
| MA-SAC | 3.45 | 13.30 | 0.1178 | 76.47 |
| **Proposed** | 0 | 11.66 | 0.0989 | 82.36 |

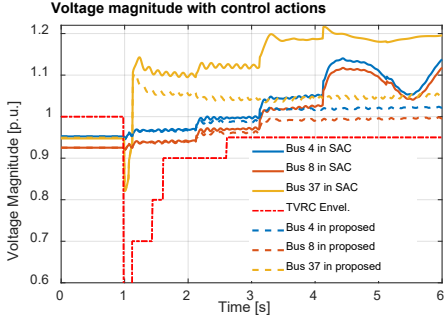

Fig.4. Comparison of voltage profiles with the MA-SAC method in a random scenario on buses 4, 8, and 37

The proposed algorithm has an average voltage deviation of about 0.0989 p.u., with a higher success rate $R_{TVRC}$. This method largely restores the severe voltage dips, compared with the case if no control is applied, which has a large average deviation of 0.1883 p.u. below the TVRC standard. It verifies that the proposed algorithm is capable of lifting the nodal voltages to the TVRC standards in a highly efficient manner. Moreover, the MA-DQN method provides the actions that result in more TVRC voltage violations and higher load costs, and even the voltage instability issue, since it lacks the adaptivity to varying operation and scalability in a large action-state space. Compared with the MA-SAC method, the proposed algorithm further displays the superiority of coordinated control, represented by a lower proportion of load loss (11.66% vs. 13.30%). This is because the MA-SAC without the attention mechanism might produce suboptimal control policies, whilst the coordination among the agents is not fully explored. Therefore, the control performance and economic benefits of online UVLS are largely improved by the proposed algorithm.

*3) Illustrative Case.* Without loss of generality, we choose a test case randomly, in which a line-to-line fault happens at line 33-34, to illustrate. Fig. 4 shows the voltage trajectories with the time-series control strategies from the proposed algorithm and MA-SAC on buses 4, 8, and 37. Through the proposed method, the voltages at the buses recover quickly above the TVRC envelope. In contrast, the actions from the MA-SAC method shed more loads and result in overvoltage and voltage instability issues. The actions from both methods decide to shed loads on buses 37 and 41, which have closer electrical distances from the unknown fault location than other controllable nodes. In this case, the DRL agents adaptively decide other controllable nodes not to attend to the control. Moreover, the proposed method sheds 20% less loads on buses 37 and 41, resulting in much lower shedding costs.

*4) Computation Time.* During the online test, the average decision-making time in each agent is 0.21 milliseconds per action step. Hence, it is incredibly promising to apply the proposed algorithm to real-time UVLS for power system emergency control.

## V. CONCLUSION

This paper proposes a cooperative MADRL-based UVLS algorithm for decentralized emergency voltage control. The high adaptivity of the proposed UVLS framework is realized by strategically combining the merits of the discrete-actor-critic DRL structure and a cutting-edge attention mechanism together. Case study verifies the merits of the proposed decentralized method over several MADRL-based algorithms.